\documentclass[%
 reprint,
 amsmath,amssymb,
 aps,
prb,
]{revtex4-1}

\usepackage{graphicx}
\usepackage{dcolumn}
\usepackage{bm}
\usepackage{hyperref}

\begin{document}

\preprint{APS/123-QED}

\title{Formation of spin droplet at {$\nu=5/2$} in an asymmetric quantum dot under quantum Hall conditions}

\author{H. Atci$^1$$^,$$^2$ and A. Siddiki$^3$}
\affiliation{$^1$ ETH Zurich, Department of Physics, CH-8093 Zurich, Switzerland}
\email{huseyinatci@gmail.com}
\affiliation{$^2$ Istanbul University, Physics Department, 34134 Vezneciler-Fatih, Istanbul, Turkey}
\affiliation{$^3$ Mimar Sinan Fine Arts University, Physics Department, 34380 Bomonti-Sisli, Istanbul, Turkey}

\date{\today}

\begin{abstract}
In this work, a quantum dot that is defined asymmetrically by electrostatic means induced on a GaAs/AlGaAs heterostructure is investigated to unravel the effect of geometric constrains on the formation of spin droplets under quantised Hall conditions. The incompressibility of exciting $\nu=5/2$ state is explored by solving the Schr\"{o}dinger equation within spin density functional theory, where the confinement potential is obtained self-consistently utilising the Thomas-Fermi approximation. Our numerical investigations show that the spatial distribution of the $\nu=2$ incompressible strips and electron occupation in the second lowest Landau level considerably differ from the results of the laterally symmetric quantum dots. Our findings yield two important consequences, first the incompressibility of the intriguing $\nu=5/2$ state is strongly affected by the asymmetry, and second, since the Aharonov-Bohm interference patterns depend on the velocity of the particles, asymmetry yields an additional parameter to adjust the oscillation period, which imposes a boundary condition dependency in observing quasi-particle phases. 
\begin{description}
\item[PACS numbers]
73.43.Cd, 73.21.La
\end{description}
\end{abstract}

\pacs{Valid PACS appear here}
\maketitle


\section{Introduction}
The fractional quantum Hall states assuming even-denominator Landau level (LL) filling factor\cite{Willett1987:1776} have recently come into the limelight both in the experimental\cite{Dean2008:146803, Choi2008:081301(R), Ying2010:176807, Shabani2010:246805, Pan2011:206806} and theoretical investigations\cite{Peterson2008:155308, Peterson2008:016807, Wojs2010:096802} due to a theoretical prediction that they are promising candidates to be useful for topological quantum computing\cite{Nayak2008:1083}. The use of the topological charge is constrained by the measurements, which should be able to readout the qubits. Hence, interferometers are indispensable for the implementation of topological quantum computation. The topological quantum computation proposal seeks to explore quasiparticle statistics of a particular fractional quantum Hall state, namely filling factor $\nu=5/2$, which is believed to obey non-abelian statistics\cite{Moore1991:362}. Theoretical intellections based on the analysis of composite fermions form a p-wave paired state, described by Moore-Read-Pfaffian wave function being a trial wave function for the ground state at filling factor $\nu=5/2$. Experimental investigation of the non-Abelian statistics of the $\nu=5/2$ fractional quantum Hall state can be performed by Fabry-Perot interferometry (FPI) of quasiparticles, leaning on the Aharonov-Bohm (AB) phase\cite{Sarma2005:166802, Stern2006:016802}. Recently, there have been advances in realising quantum Hall edge-state based interferometers at fractional filling factors at the lowest LL\cite{Bishara2008:165302, Baer2013:023035}. Much of the attention has been focused on the filling factor $\nu=5/2$ state in quantum dots (QDs) which consist of the two LLs\cite{Rasanen2008:041302(R), Atci2013:155604, Rosenow2009:155305}. For an infinite system the lowest LL is fully occupied and is spin compensated, whereas the second lowest LL is spin polarised. For a finite system, e.g. a QD, the formation of spin droplet (SD) is expected, due to competition between the confinement potential and interactions. Once the spin polarised electrons in the second lowest LL is localised to the centre of the QD emanating from exchange-correlation effects, they are called SD, which is a many-body phenomenon of interacting electrons\cite{Harju2006:126805, Rasanen2008:041302(R), Saarikoski2008:195321}. 

Rasanen et al.\cite{Rasanen2008:041302(R)} show that theoretical evidence of SD formation in large ($N\geqslant30$) QDs at the filling factor range $2<\nu<3$ by using numerical many electron methods. Their calculations indicate that the paired electron state breaks down leading to fragmentation of spin and charge densities in parabolic external confining potentials. They point out that evidence of the fragmentation can be tested by investigating the spatial dependence of the spin and charge densities in different geometries. Before our previous work\cite{Atci2013:155604}, there has been no systematic investigation on the stability of SDs at the filling factor range $2<\nu<3$ considering broken rotational symmetry. There it is shown that, the broken rotational symmetry does not considerably affect the stability of SD formation. However, the stability of a SD under complete break down of both the axial and the rotational symmetry of the confinement is still under debate and such an asymmetric confinement is much more realistic in investigating experimental systems. Here, we perform numerical investigation of the formation of SDs in a half-side etched, half-side gated QD yielding an asymmetric confining potential. 

\begin{figure}
{\centering
\includegraphics[width=.8\linewidth]{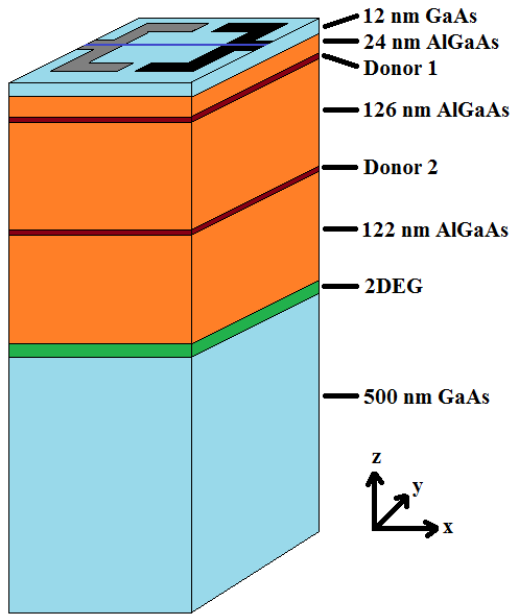}
\caption{(Color online) \label{fig:1} Sketch of the layer sequence of GaAs/AlGaAs heterostructure, which has the dimensions of $2550\times2550\times784$ nm$^{3}$. The heterostructure is grown on a thick GaAs substrate, where 2DEG is formed at the interface of the GaAs/AlGaAs denoted by green region. Silicon donor layers (dark brown regions), distributed homogeneously on the $xy$ plane, provide electrons both to surface and to the 2DEG. The light gray and black regions on the surface are shown as the etched and the metallic gate regions, respectively.}}
\end{figure}

The outline of this paper is organized as follows. In Sec.~\ref{sec:level2} we briefly describe the GaAs/AlGaAs heterostructure and introduce confinement potential of the device with a self-consistent electronic calculation. Then we solve computationally the many-electron problem using the spin density functional theory (SDFT). In Sec.~\ref{sec:level3}, we analyze the electronic structure of fragmented quantum Hall states and show that  the incompressibility of the $\nu=5/2$ state is affected by the asymmetry. In Sec.~\ref{sec:level4} concludes our work with discussion of the relevance of our findings with formation of the spin droplet at filling factor $\nu=5/2$ quantum Hall state in confined two dimensional electron gas (2DEG). The paper is summarized in Sec.~\ref{sec:level5}.


\section{\label{sec:level2}The Geometry and Model Hamiltonian}

\begin{figure}
{\centering
\includegraphics[width=1.\linewidth]{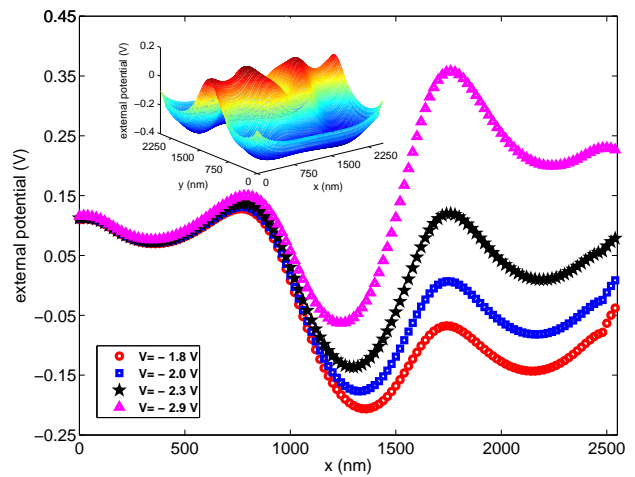}
\caption{(Color online) \label{fig:2} The confinement potential for electrons at the interface of the GaAs/AlGaAs heterostructure obtained with self consistent electronic calculations. Inset graph shows the symmetric confinement potential (when applied -2.3 V) more clearly.}}
\end{figure}

The system we study is shown in Figure~\ref{fig:1},a GaAs/AlGaAs heterostructure\cite{Camino2005:155313, Camino2007:076805} consists of two $\delta$-doped silicon donor layers (dark brown regions) which provide electrons to the 2DEG (green region) forming 284 nm below the surface at the interface between GaAs and AlGaAs. The donor layers lie 122 nm and 248 nm above the 2DEG and have the surface densities $2.5\times10^{15}$ m$^{-2}$ and $1.7\times10^{16}$ m$^{-2}$, respectively. The physical dimensions of heterostructure are taken as $L_{x} = L_{y} = 2550$ nm via a matrix of $128 \times 128$ mesh points. The left side of the heterostructure (light gray region) is etched 80 nm below the surface and different voltages are applied to metallic gate which is deposited on the surface of heterostructure at the right side (black region). Metallic gate is biased with -1.8 V, -2.0 V, -2.3 V and -2.9 V, respectively. The realistic modelling of the 2DEG to be located in the $z=0$ relies on solving the Poisson equation in three dimensions self-consistently within the Thomas-Fermi approximation (TFA) which describes realistically the electronic distribution for the given boundary conditions, set by the GaAs/AlGaAs heterostructure and surface patterns. To calculate electron and potential profiles within the TFA the computational effort is much simpler than other quantum mechanical calculations and yields compatible results. The spatial distrubition of the electron density is calculated within the TFA\cite{Siddiki2003:125315, Siddiki2004:195335},
\begin{equation}
n_{\rm el}(x,y)=\int D(E)f[E+V_{\rm tot}(x,y)-\mu^{*}] dE 
\end{equation}
with $D(E)\equiv \sum \delta (E-E_{\rm n})$ describing the local density of states (LDOS), $f(E)=1/[exp(E/k_{\rm B}T)+1]$ as Fermi function, $\mu^{*}$ as the electrochemical potential, $k_{\rm B}$ as the Boltzmann constant, and $T$ as the temperature. We write the total potential energy of an electron as
\begin{equation}
V_{\rm tot}(x,y)=V_{\rm ext}(x,y)+V_{\rm H}(x,y)
\end{equation}
where $V_{\rm ext}(x,y)$ and $V_{\rm H}(x,y)$ are the external (confining) potential composed of gates and donors and the electron-electron interaction (Hartree) potential, respectively. Since the Hartree potential depends on the electron density via\cite{Oh1997:13519}
\begin{equation}
V_{\rm H}(x,y)=\dfrac{2e^{2}}{\epsilon}K(x,y,x^{'},y^{'})n_{\rm el}(x^{'},y^{'})dx^{'}dy^{'},
\end{equation}
where $-e$ is the electron charge, $\epsilon$ is the dielectric constant (=12.4 for GaAs) and the kernel $K(x,y,x^{'},y^{'})$ is the solution of the 2D Poisson equation with appropriate boundary conditions, Eqs. (1) and (2) complete the self-consistent loop\cite{Guven2003:115327}, which can be solved by a numerical iteration. This kernel can be found in a well-known textbook\cite{Morse1953:1240}. To solve the Poisson equation we use a code, developed by A. Weischelbaum, which is succesfully applied in earlier studies\cite{Arslan2008:125423, Mese2010:354017, Weichselbaum2003:056707, Weichselbaum2006:085318}. Overall the code provides a reliable description of the potential landscape both in the absence and presence of electron-electron interactions in the 2DEG. To obtain the confining potential, we use half-side etched and half-side gated heterostructure depicted in Fig ~\ref{fig:1}. In Fig. ~\ref{fig:2}, we show results of the confining potential that is nearly symmetric when -2.3 V is applied to metallic gate, however the spatial symmetry is lifted for gate voltages -1.8 V, -2.0 V and -2.9 V, respectively. 


Our investigation focuses on QDs, which are induced on GaAs/AlGaAs heterostructures confined to a 2D plane and most importantly the electrostatic confinement is spatially asymmetric, under quantum Hall conditions. We use the effective-mass approximation with considering material parameters of a GaAs semiconductor medium, i.e. the effective mass $m^{*}=0.067m_{\rm e}$ and the dielectric constant $\epsilon=12.4$. The $N$-electron system in an external confining potential and magnetic field is described by an effective-mass Hamiltonian 
\begin{multline}
H=\dfrac{1}{2 m^*}\sum^N_{i=1}\Big[-i\hbar \nabla_i + e{\bf A}({\bf r}_{\rm i})\Big]^2 + \sum^N_{i<j}\frac{e^2}{4\pi\epsilon_0\epsilon |r_{\rm i}-r_{\rm j}|}\\
+\sum^N_{i=1}\Big[V_{\rm ext}(r_{i})+g^{*}\mu_{\rm B}BS_{\rm z,\rm i}\Big],
\end{multline}
where $N$ defines the total electron number inside the quantum dot, $\textbf{A}=B/2(-y,x,0)$ is the vector potential given in the symmetric gauge for the homogeneous magnetic field $\textbf{B}=B\widehat{z}$ perpendicular to the plane and $V_{ext}(\textbf{r})$ is the external confining potential in the $xy$ plane (see Figure ~\ref{fig:2}). The last term is the Zeeman energy arising from the application of an external magnetic field with electron spin. Here, $g^{*}=-0.44$ is the effective gyromagnetic ratio, $\mu_{B}=e\hbar/2m_{\rm e}$ is the Bohr magneton, and $S_{\rm z}=\pm\frac{1}{2}$ represents the up and down spins, respectively. We solve the Schr\"{o}dinger equation, $H\Psi=E\Psi$, associated with the $N$-electron Hamiltonian in Eq.(4) using numerical approaches, namely the SDFT in the self-consistent Kohn-Sham formulation\cite{Kohn1965:A1133}. To obtain ground state energy depends on spin densities $\sigma=n_{\uparrow}(\textbf{r}),n_{\downarrow}(\textbf{r})$ of a system of interacting electrons, the Kohn-Sham states are solved from the Kohn-Sham equation\cite{Barth1972:1629}
\begin{equation}
\biggr[T_{\rm 0}^{\sigma}+ V_{\rm KS}^{\sigma}(\textbf{r})\biggr]\varphi_{\rm i\sigma}(\textbf{r})=E_{\rm i\sigma}\varphi_{\rm i\sigma}(\textbf{r}).
\end{equation}
Here, the first term is the kinetic energy functional of noninteracting electrons with spin densities and the second term is Kohn-Sham potential $V_{\rm KS}^{\sigma}(\textbf{r})$, defined as
\begin{equation}
V_{\rm KS}^{\sigma}(\textbf{r})=V_{\rm ext}^{\sigma}(\textbf{r})+V_{\rm H}^{\sigma}(\textbf{r})+V_{\rm xc}^{\sigma}(\textbf{r}),
\end{equation}
where the sum of the external confining potential $V_{\rm ext}^{\sigma}(\textbf{r})$ acting on the interacting system, the classic electrostatic or Hartree potential $V_{\rm H}^{\sigma}(\textbf{r})$ and the exchange-correlation potential is given by
\begin{equation}
V_{\rm xc}^{\sigma}(\textbf{r})=\dfrac{\delta E_{\rm xc}^{LSDA}}{\delta n^{\sigma}(\textbf{r})}=\int d \textbf{r}  n(\textbf{r}) e_{\rm xc}(n(\textbf{r}),\zeta(\textbf{r}))
\end{equation}
where $e_{\rm xc}$ is the exchange-correlation energy per electron depends on the total spin density $n=n_{\uparrow}+n_{\downarrow}$ and spin polarization $\zeta=(n_{\rm \uparrow }-n_{\rm \downarrow}) /n$. To calculate exchange-correlation energy $E_{\rm xc}$, we use the local spin density approximation (LSDA) with a parametrization provided by Attaccalite \textit{et al.}\cite{Attaccalite2002:256601}. The SDFT scheme together with LSDA leads to good numerical accuracy and produce reliable results in comparision with quantum Monte Carlo calculations in quantum dot systems\cite{Rasanen2008:041302(R), Saarikoski2008:195321, Rogge2010:046802}. In the SDFT calculations, we utilise the OCTOPUS\cite{Rubio2003:60-78, Castro2006:2465} code package (published under the General Public License) built on the real space grid discretization method which allows realistic modeling of two dimensional systems. Related technical details can be found in References (36) and (37). To solve Schr\"{o}dinger equation the conjugated gradient algorithm is used. 

We point out that for the $\nu\geq1$ regime the filling factor can be defined $ \nu = 2N / N_{\rm 0LL} $, approximately in the QDs. Here, $N_{\rm 0LL} $ is the number of electrons in the lowest LL. Theoretical investigation of many-body effects in a realistic QD involves the self-consistent solution of the Schr\"{o}dinger equation. To address the many body problem, we use SDFT with LSDA\cite{DG}. 

\begin{figure}
{\centering
\includegraphics[width=0.8\linewidth]{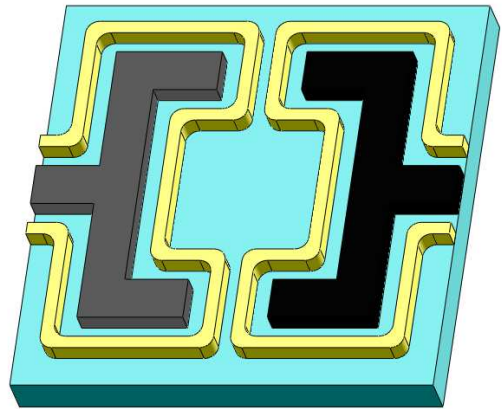}
\caption{(Color online) \label{fig:3} A FPI device. Current flows along the counterpropagating edge states (shown by yellow lines). Tunnelling occurs in the two narrow constrictions, when the edge channels are at a sufficient proximity.}}
\end{figure}


\section{\label{sec:level3}Incompressibility}

\begin{figure}
{\centering
\includegraphics[width=1.\linewidth]{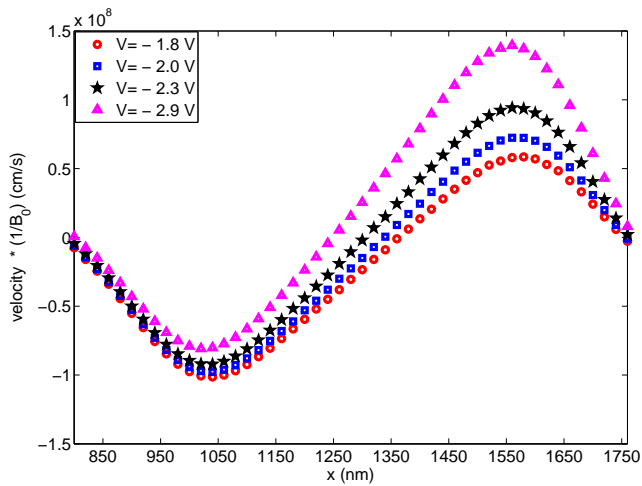}
\caption{(Color online) \label{fig:4} The slopes of the confinement potential of the sample gives the electron velocity as a function of the position.}}
\end{figure}

Interference phenomena in the quantum Hall regime characterized by strong electron-electron interactions is a very prominent topic in transport\cite{Camino2005:155313, Camino2005:075342, Ji2003:415}. To investigate interference effects at the edge of the quantum Hall system a type of FPI can be adressed. The novel geometry called FPI which shows edge channels (yellow lines) is shown Fig. ~\ref{fig:3}. Interference occur when current-carrying two edge channels are at close proximity allowing scattering to provide partitioning. The current-carrying edge channels in the FPI acquire a phase determined by the AB effect and the number of quasiparticles. This resultant phase arises from different velocities of the quasi-particles on different paths and can be controlled by either changing the magnetic field or the area of the interferometer\cite{McClure2009:206806}. The phase difference is strongly affected by the electron velocity and is an important transport parameter for interferometers. To determine the edge channel velocity in the quantum Hall regime McClure et al.\cite{McClure2009:206806} used interference to explain checkerboard patterns at a FPI geometry. Using the gradient of the confining potential it is straight forward to obtain electron velocities within a self-consistent screening theory. Fig.~\ref{fig:4} shows electron velocities that flow through edge channels which are at the order of $10^7$ cm/s. This result is consistent with previous theoretical calculations\cite{Eksi2007:075334}.

Self-consistent screening calculations depends on the electron-electron interactions explained by the recent theoretical works show that the 2DEG contains two different kinds of regions: the compressible and incompressible regions\cite{Siddiki2007:045325, Salman2012:1405}. In the compressible region a partially filled LL with its high density of states is pinned to the Fermi energy and the electronic system behaves like a quasi-metal and screens completely the confining potential. In contrast to compressible region, when the Fermi level is between two consequent LL, electrons do not contribute to screening locally i.e. the confining potential could not be screened perfectly. Here, the system presents a constant electron density and is called to be incompressible, behaving like a quasi-insulator. 

In a 2DEG the incompressibility is defined as 
\begin{equation}
\kappa^{-1}=-S\biggr(\frac{\partial P}{\partial S}\biggr)_{N}=S\biggr(\frac{\partial^{2}E}{\partial S^{2}}\biggr)_{N},
\end{equation}
where the pressure $P=({\partial E}/{\partial S})$ is the change of energy according to the area change, $S$ is the area of the 2DEG, $N=S n_{\rm el}$ is the total number of the electrons, and the total energy of the system is $N$ times the ground state energy per particle, $E_{\rm tot}=N\varepsilon(n_{\rm el})$. Using standart manipulations, we may rewrite incompressibility according to changing the chemical potential instead of changing the electron number density as
\begin{equation}
\kappa^{-1}={n_{\rm el}^2}\biggr(\frac{d\mu}{dn_{\rm el}}\biggr),
\end{equation}
where the chemical potential $\mu$ is related to the total energy $E_{\rm tot}$ by
\begin{equation}
\mu=\frac{\partial E_{\rm tot}}{\partial N}=\frac{\partial (E_{\rm tot}/S)}{\partial n_{\rm el}}.
\end{equation}
From Eq.(9) we can tell the system is incompressible if the chemical potential increases discontinuously as a function of density. The incompressibility of a 2DEG is a fundamental thermodynamic quantity and proportional to the thermodynamic density of states (TDOS), $D_{\rm T}(\mu,B)=dn_{\rm el}/d\mu$, which is the rate of change of the chemical potential with electron concentration. To illustrate the meaning of the term thermodynamic density of states, we calculate 
\begin{equation}
\frac{dn_{\rm el}}{d\mu}=\frac{d}{d\mu}\int dE D(E)f(E)=\int dE D(E)\frac{df(E)}{d\mu}.
\end{equation}
The TDOS tells us how much the ground state energy changes when an additional particle is added to the system\cite{Ezawa}.

\begin{figure}
{\centering
\includegraphics[width=1.\linewidth]{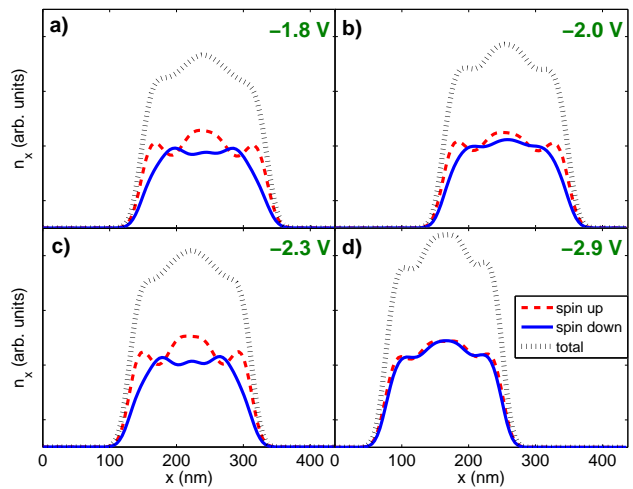}
\caption{(Color online) \label{fig:5} Spin-up, spin-down and total electron density of quantum Hall state in a QD includes 30-electrons calculated with the SDFT at $\nu=5/2$ for various gate voltages.}}
\end{figure}


\section{\label{sec:level4}Results}
Utilizing the confinement potential obtained via solving the Poisson equation in 3D, we calculate the corresponding electron density distribution within the QD. Fig~\ref{fig:5}. depicts the spin densities for a QD that contains 30 electrons at $\nu=5/2$ state. Our numerical investigations show that the position of the incompressible strips and electron numbers in the second lowest LL differ from the results of the symmetric quantum Hall devices. Although the position of the incompressible strip shifts to the right side while appling -1.8 V and -2.0 V to the metallic gate, it shifts to the left side for -2.9 V. Even so, it looks nearly symmetric when metallic gate is biased with -2.3 V. The electron numbers in the SD are 3, 4, 4, 5 for -1.8 V, -2.0 V, -2.3 V, -2.9 V, respectively.
\begin{figure}
{\centering
\includegraphics[width=1.\linewidth]{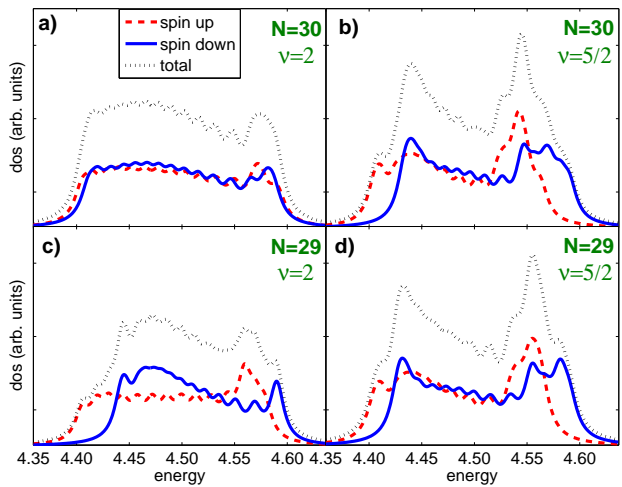}
\caption{(Color online) \label{fig:6} LDOS for different electrons numbers and filling factors.}}
\end{figure}

To check spin dependency of spin droplet we obtain local density of states (LDOS) when total electron number is $N=30$ and $N=29$ inside the quantum dot for filling factor $\nu=2$ and $\nu=5/2$. Fig.~\ref{fig:6} shows LDOS for $\nu=2$ and $\nu=5/2$. We see that there is a peak in both 30-electrons and 29-electrons for $\nu=5/2$ and this indicates presence of the spin droplet, incompressible droplets of spin polarized of second Landau level (SLL) electrons. The spin splitting of the SLL is analogous to the Stoner criterion\cite{Stoner1938:372}, which states in the presence of correlations between electrons with same spin and high density of states near the Fermi level. Now, the system favors ferromagnetic alignment that reduces the degeneracy.

\begin{figure}
{\centering
\includegraphics[width=1.\linewidth]{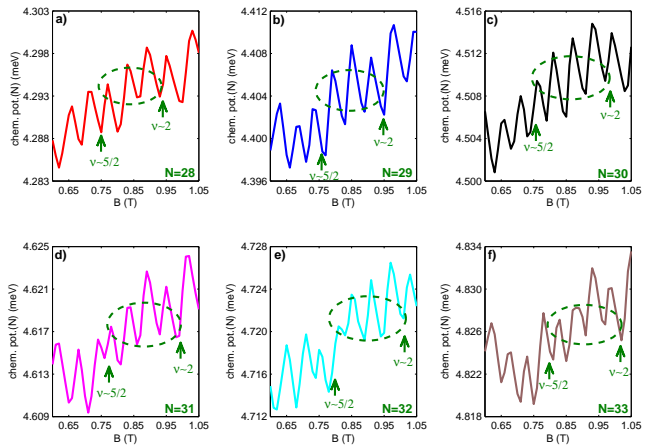}
\caption{(Color online) \label{fig:7} Chemical potentials obtained symmetric confining potential for various $N$-electron symmetric quantum dots as a function of the magnetic field $B$. The dotted ellipse denots the spin-droplet regime.}}
\end{figure}

\begin{figure}
{\centering
\includegraphics[width=1.\linewidth]{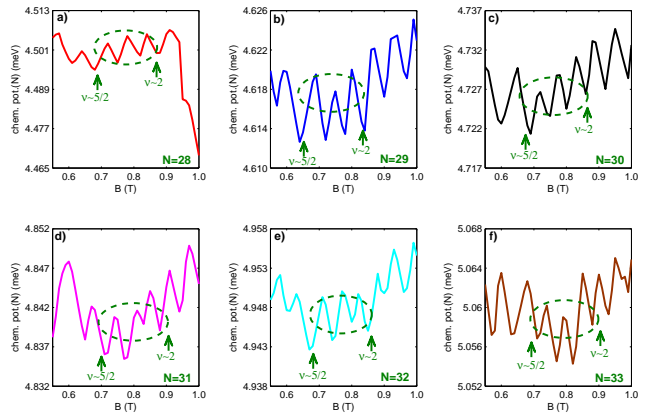}
\caption{(Color online) \label{fig:8} Chemical potentials obtained asymmetric confining potential for various $N$-electron symmetric quantum dots as a function of the magnetic field $B$. The dotted ellipse denots the spin-droplet regime.}}
\end{figure}

We discuss the consistence of finite size counterparts of the integer and fractional quantum Hall states in the quantum dot which gives characteristic properties in the chemical potential. The existed and the properties of these states can be defined by chemical potential $\mu(N,B)=E_{\rm tot}(N,B)-E_{\rm tot}(N-1,B)$,which is the energy needed to add the $N^{th}$ electron in the system of $N-1$ electrons. Fig ~\ref{fig:7}. and Fig ~\ref{fig:8}. and  shows results for chemical potentials in the comparison with various electron numbers for symmetric and asymmetric potentials, respectively. Sudden jumps in oscillations correspond to filling factor $\nu=5/2$. The observed jumps clearly indicates that the 5/2 state is compressible, once again regardless of the symmetry of the confinement.

\section{\label{sec:level5}Summary}

In conclusion, we have seen the electronic compressibility of a 2DEG in the fractional quantum Hall regime. The compressibility images show quasi-insulating, incompressible, strips that seperate region of electron liquid of near integer filling. The incompressible strips form near the boundary of the sample due to a smooth density gradient. A potential step accompanies the incompressible strip. We see that the incompressibility of the filling factor $\nu=5/2$ is strongly affected by the asymmetry of the potential of the sample. These findings are similar to previous observation, even if the confinement potential is no longer asymmetric. Such a stable configuration of the spin droplet state also enables us to claim that this state is universal. Our numerical investigations also show that the position of the incompressible strips and electron numbers in the SLL differ from the results of the symmetric quantum Hall devices. Our calculations indicate that the paired electron state breaks down leading to fragmentation of spin densities. We find evidence of fragmentation in sevaral calculations but point out that our results can be tested by direct measurements of the spatial dependence of spin densities in different geometries and experimental setups. Asymmetric slope gives electron velocity and an electron with the order of $10^7$ cm/s passes through interferometers. This result is consistent with previous theoretical calculations. It is concluded that the electron velocity is an important transport parameter for the interferometers. 

\begin{acknowledgments}
The authors would like to acknowledge the Scientific and Technical Research Council of Turkey (TUBITAK) for supporting under grant no 112T264 and Marmaris Institute of Theoretical and Applied Physics (ITAP). H. A. also acknowledges support from TUBITAK under grant 2214/A.
\end{acknowledgments}


\end{document}